\documentclass[conference]{IEEEtran}

\usepackage[english, ngerman]{babel}
\usepackage[T1]{fontenc}
\usepackage[utf8]{inputenc}
\usepackage{graphicx}
\usepackage{subfigure}
\usepackage[nospace]{cite}
\usepackage{url}
\usepackage{paralist}

\usepackage{url}
\makeatletter
\def\ps@IEEEtitlepagestyle{
	\def\@oddfoot{\mycopyrightnotice \thepage}
	\def\@evenfoot{}
}
\def\mycopyrightnotice{
	{\footnotesize
		\begin{minipage}{\textwidth}
			\centering
			\textcopyright~2015 IEEE. Personal use of this material is permitted.
			Permission from IEEE must be obtained for all other uses, in any current or future media, including reprinting/republishing this material for advertising or promotional purposes, creating new collective works, for resale or redistribution to servers or lists, or reuse of any copyrighted component of this work in other works. 
			{\tt DOI:} \url{https://doi.org/10.1109/SASOW.2015.27}
		\end{minipage}
	}
}
\usepackage[hidelinks]{hyperref}
\hypersetup{
	bookmarks=true,         
	unicode=false,          
	pdftoolbar=true,        
	pdfmenubar=true,        
	pdffitwindow=false,     
	pdftitle={A Testing Scheme for Self-Adaptive Software Systems with Architectural Runtime Models},
	pdfauthor={Joachim Hänsel, Thomas Vogel and Holger Giese},
	pdfsubject={},
	pdfcreator={},
	pdfproducer={},
	pdfkeywords={},
	pdfnewwindow=true,
}

\begin{document}
\selectlanguage{english}

\title{A Testing Scheme for Self-Adaptive Software Systems with Architectural Runtime Models}

\author{\IEEEauthorblockN{Joachim Hänsel, Thomas Vogel and Holger Giese}
\IEEEauthorblockA{Hasso Plattner Institute for Software Systems Engineering\\
at the University of Potsdam, Potsdam, Germany\\
EMail: {\tt[Joachim.Haensel|Thomas.Vogel|Holger.Giese]@hpi.de}}
}

\maketitle
\pagestyle{plain}
\IEEEpeerreviewmaketitle

\begin{abstract}
Self-adaptive software systems (SASS) are equipped with feedback loops to adapt autonomously to changes of the software or environment. In established fields, such as embedded software, sophisticated approaches have been developed to systematically study feedback loops early during the development. In order to cover the particularities of feedback, techniques like one-way and in-the-loop simulation and testing have been included. However, a related approach to systematically test SASS is currently lacking. In this paper we therefore propose a systematic testing scheme for SASS that allows engineers to test the feedback loops early in the development by exploiting architectural runtime models. These models that are available early in the development are commonly used by the activities of a feedback loop at runtime and they provide a suitable high-level abstraction to describe test inputs as well as expected test results. We further outline our ideas with some initial evaluation results by means of a small case study.
\end{abstract}

\section{Introduction}
\noindent
Traditionally, software development follows an open-loop structure that requires human supervision when software systems are exposed to changing environments~\cite{Salehie&Tahvildari2009}. To reduce human supervision, software systems are equipped with feedback loops to adapt autonomously to changing environments. Such closed-loop systems are designated as self-adaptive software systems (SASS)~\cite{SEfSAS2-ROADMAP} and they are often split in two parts, an \textit{adaptation engine} realizing the feedback loops and controlling the \textit{adaptable software}~\cite{Salehie&Tahvildari2009}. As pointed out by Calinescu~\cite{Calinescu2013}, such systems will become important for safety-critical applications, where they have to fulfill high-quality standards.

Testing is an established technique for ensuring quality in traditional systems, even safety-critical ones~\cite{Nair2014689}, and processes for testing such systems exist. For instance, embedded software with its feedback loops is often systematically tested in three stages~\cite[pp. 193--208]{broekman2003testing}:%
\begin{inparaenum}[i)]
\item a simulation stage that tests the models (specification) of the software under development in a simulated or real-life environment,
\item a prototyping stage that tests the real software in a simulated environment, and finally,
\item pre-production stage that tests the real software in the real environment.
\end{inparaenum}
With each stage the software is more and more refined to the final product while testing continuously provides assurances for the software and particularly early in the development.

However, a similar systematic testing process providing continuous and early assurances does not exist for SASS. In contrast, models for substituting the environment or parts of a SASS usually cannot be obtained easily and therefore, a generic simulation environment for SASS does not exists. Consequently, testing SASS typically requires that the implementations of the feedback loops and adaptable software with its sensors and effectors are available. This impedes testing early in the development and makes it costly to remove faults in the feedback loops discovered late in the development.

Furthermore, approaches used for traditional systems are not as easily applicable to SASS as the interface between the adaptation engine and the adaptable software is often quite different from that of embedded software. SASS are usually not restricted to observing and adjusting parameters but additionally monitor and adapt the architecture of the software~\cite{Vogel-ICAC09,GarCHSS04,Kramer&Magee2007}, thus requiring support of structural adaptations~\cite{McKinley+2004}.

Some approaches address the testing of SASS but only for later development stages~\cite{Goldsby2008,Zhao2006,Eberhardinger2014,Fredericks2015,Fredericks2014a} when the systems have already been deployed. Others do promote testing in earlier stages but they still assume an executable and complete SASS to run the tests against~\cite{Pueschel2014,Wang2007,Camara2014}. Testing of only parts of the feedback loop is not supported. In contrast, we consider testing parts of a SASS as a precondition to early validation since a system with the completely implemented adaptable software and feedback loop is only available in the latest development stages.      

Therefore, we propose a systematic testing scheme for SASS that allows engineers to test the feedback loops (adaptation behavior) early in the development by exploiting runtime models. Such models represent the adaptable software and environment and they are typically used at runtime to drive the adaptation~\cite{MC.2009.326}. Our approach leverages early testing of SASS by using \textit{architectural} runtime models that are available early in the development and are commonly used by the activities of a feedback loop. Therefore, feedback loop activities such as monitor, analyze, plan, and execute (cf.~\mbox{MAPE-K}~\cite{Kephart2003}) can be individually tested while the whole feedback loop and the adaptable software do not have to be implemented yet. In contrast, the non-implemented parts are simulated based on the runtime models. 
Consequently, the feedback loop can be modularly tested while the different parts of the loop can be incrementally refined and implemented until they replace the simulated parts. Moreover, we expect reduced costs of testing since we do not require final or experimental implementations of certain feedback loops parts to test other parts.

The rest of the paper is structured as follows. We describe preliminaries in Section~\ref{sec:preliminaries} and the benefits of runtime models for testing in Section~\ref{sec:runtime-models}. Then, we discuss our approach by means of one-way (Section~\ref{sec:one-way-testing}), in-the-loop  (Section~\ref{sec:in-the-loop-testing}), and online (Section~\ref{sec:online-testing}) testing. Finally, we sketch an initial evaluation in Section~\ref{sec:evaluation}, contrast our approach with related work in Section~\ref{sec:related-work}, and conclude the paper in Section~\ref{sec:conclusion}.

\section{Preliminaries}
\label{sec:preliminaries}
\noindent
In this section we discuss preliminaries of the presented testing: MAPE-K feedback loops and architectural RTM.

\subsection{MAPE-K Feedback Loops}
\begin{figure}[h]
\centering
\includegraphics[width=0.65\columnwidth]{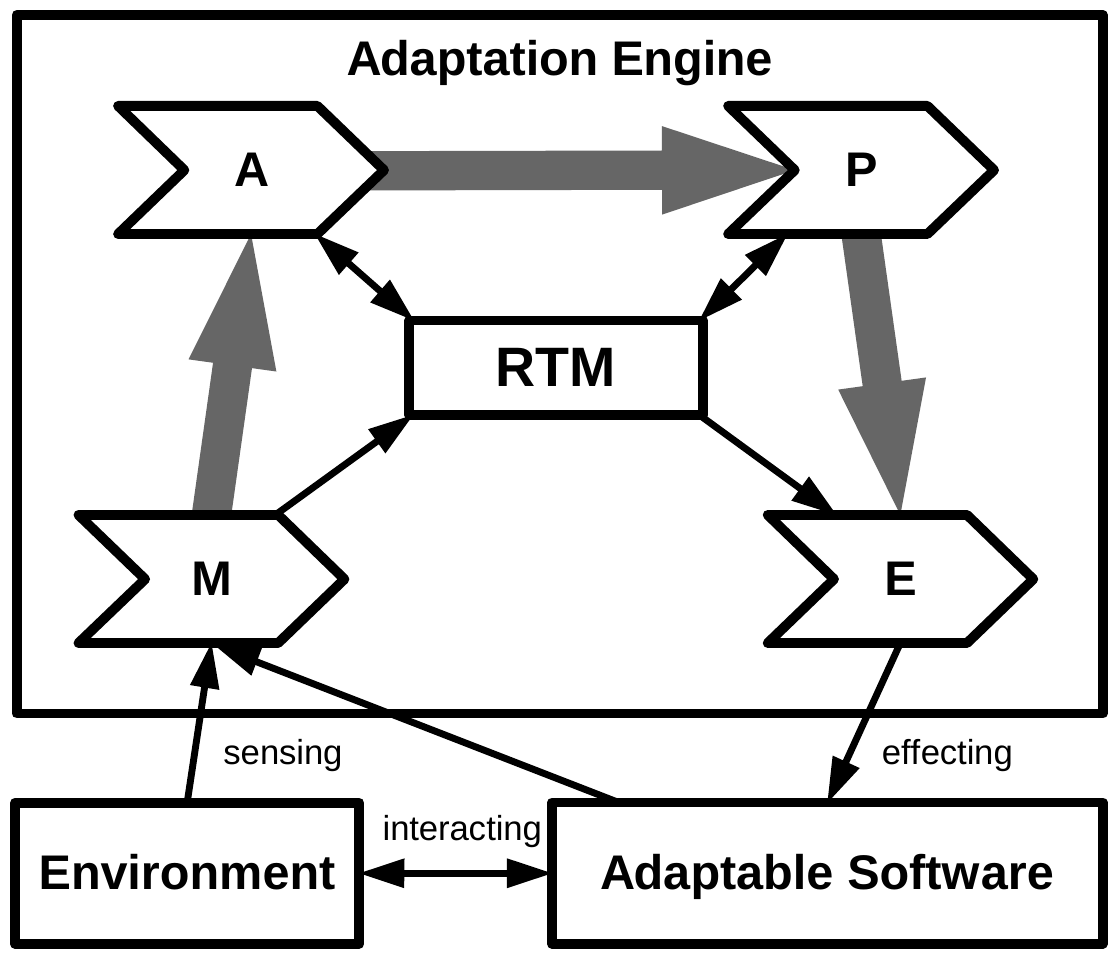}
\caption{MAPE-K Feedback Loop with a Runtime Model (RTM).}
\label{fig:mapeks}
\end{figure}

\noindent
The development of SASS typically follows the \textit{external approach}~\cite{Salehie&Tahvildari2009} that separates adaptation from domain concerns by splitting up the software in two parts: an \textit{Adaptation Engine} for the adaptation concerns and an \textit{Adaptable Software} for the domain concerns while the former \textit{senses} and \textit{effects} and thus, controls the latter. This constitutes a \textit{feedback loop} that realizes the self-adaptation (see Figure~\ref{fig:mapeks}). The engine senses as well the \textit{Environment} with which the adaptable software \textit{interacts}.

The resulting feedback loop between the engine and the software can be refined according to the \mbox{\textit{MAPE-K}} reference model~\cite{Kephart2003}. This model considers the activities of \textbf{M}onitoring and \textbf{A}nalyzing the software and environment and, if needed, of \textbf{P}lanning and \textbf{E}xecuting adaptations to the software. All activities share a \textbf{K}nowledge base as illustrated by a runtime model (\textit{RTM}) in Figure~\ref{fig:mapeks} and discussed in the following.

\subsection{Architectural Runtime Models}\noindent
The external approach as previously discussed requires that the adaptation engine has a representation of the adaptable software and environment to perform self-adaptation. This representation is often realized by a causally connected \textbf{R}un\textbf{t}ime \textbf{M}odel (\textit{RTM})~\cite{MC.2009.326}. A causal connection means that changes of the software or environment are reflected in the model and changes of the model are reflected in the software (but, not in the environment being a non-controllable entity).

Considering Figure~\ref{fig:mapeks}, an RTM can be used as a knowledge base on which the MAPE activities are operating. The monitor step observes the software and environment and updates the RTM accordingly. The analyze step then reasons on the RTM to identify any need for adaptation. Such a need is addressed by the plan step to prescribe an adaptation in the RTM, which is eventually enacted to the software by the execute step.

Using RTMs in self-adaptive software provides the benefits of creating appropriate abstractions of runtime phenomena that are manageable by the feedback loops and of applying automated model-driven engineering (MDE) techniques~\cite{MC.2009.326}.

The software architecture has been identified as such an appropriate abstraction level for representing the adaptable software and environment and for supporting structural adaptation~\cite{MC.2009.326,GarCHSS04,Kramer&Magee2007,Vogel-ICAC09,McKinley+2004}. Hence, \textit{architectural} RTMs of the adaptable software are used by a feedback loop to reflect on the state of the software and environment. Such state-aware models can be enriched by a feedback loop to cover, for instance, the history or time series of states and executed adaptations, which results in history-/time-aware models.

In our research on self-adaptive software such as~\cite{VG-TAAS-EUREMA}, we evaluate our work by using \textit{mRUBiS}\footnote{Modular Rice University Bidding System: \url{http://www.mdelab.de}}, an internet marketplace on which users sell or auction products, as the adaptable software. A single shop on the marketplace consists of 18 components and we may scale up the number of shops. 
For a self-healing scenario, we created architectural runtime models of mRUBiS and defined different types of failures based on the models. These failures have to be handled by the adaptation engine. Examples of such failures are exceptions emitted by components, unwanted life-cycle changes of components, the complete removal of components because of crashes, and repeated occurrences of these failures. Based on that, we experiment with different adaptation mechanisms and can also exploit the models for testing as discussed in the following.

\section{Exploiting Runtime Models for Testing}
\label{sec:runtime-models}
\noindent
In the following, we assume a SASS that follows the \mbox{MAPE-K} cycle with runtime models (RTMs) as schematically depicted in Figure~\ref{fig:mapeks}. If the RTMs are just self-aware and reflect the current state of the adaptable software and environment, we can make the following two observations:

(1) The behavior of the system can be described by a sequence of steps 
$(\to_{AS} or \to_{ENV})^* \to_M \to_A \to_P \to_E (\to_{AS} or \to_{ENV})^*; \dots$
where
$\to_{AS}$ denotes a step of the adaptable software,
$\to_{ENV}$ denotes a step of the environment,
$\to_M$ denotes the complete monitoring step,
$\to_A$ denotes the complete analysis step,
$\to_P$ denotes the complete planning step, and
$\to_E$ denotes the complete execute step.

(2) The interface between those steps can be described by different states $S_{i}$ of the RTM if we do not consider the input of the monitoring and the output of the execute step:
$(\to_{AS} or \to_{ENV})^* \to_M S_1^M \to_A S_1^A \to_P S_1^P \to_E (\to_{AS} or \to_{ENV})^*; \to_M S_2^M\dots$
where
$S^M_i$ denotes the RTM state after the $i$-th monitoring, 
$S^A_i$ the RTM state after the $i$-th analysis, and
$S^P_i$ the RTM state after the $i$-th planning.

\begin{figure}[h]
\centering
\includegraphics[width=1\columnwidth]{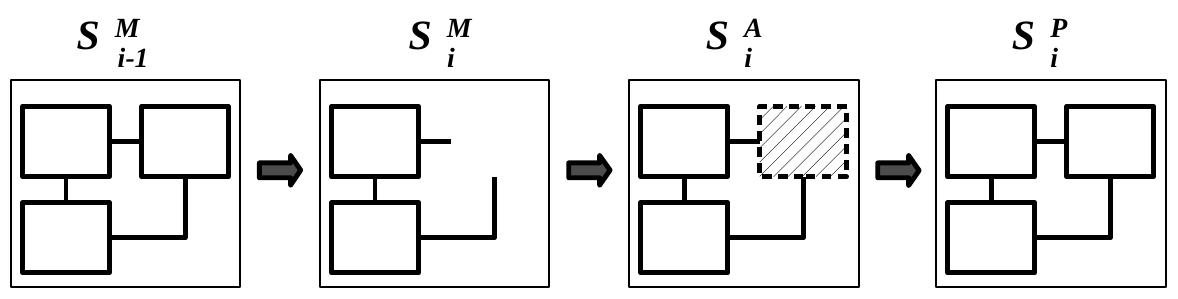}
\caption{Example Trace for a Self-Healing Scenario.}
\label{fig:trace}
\end{figure} 

Consider the self-healing example in Figure~\ref{fig:trace}. An intact architecture is monitored and results in RTM $S_{i-1}^M$. For now, analysis and planning are not required to take action since the architecture is not broken. Without an adaptation, the execute step will do nothing either. We can directly proceed with the next steps in the environment or adaptable software. Due to either an environmental influence or some failure in the adaptable software ($\to_{ENV}$ or $\to_{AS}$), a component of the architecture is removed. In the next step, this is monitored as RTM $S_{i}^M$. The result of the analysis step $\to_A$ is the annotated RTM $S_{i}^A$ that marks the missing component. The planning step $\to_P$ constructs a repaired RTM $S_{i}^P$ which will be applied to the adaptable software in the next step by $\to_E$. 

These two observations indicate that the different states of the RTM are the key element to describe the input/output behavior of the MAPE activities concerning their communication with the adaptable software. 
Moreover, the RTMs also facilitate considering the required behavior of the adaptation engine at a much higher level of abstraction than the events observed by the monitoring step and the effects triggered by the execute step.\footnote{\label{fot:ignorepar}We ignore here the case that the adaptable software and environment change while the feedback loop is running. While this case could not be excluded in general, we may neglect it due to the considered abstraction level as supported by architectural runtime models. That is, oftentimes the architecture does not change very frequently, for instance, due to failures.} 
Consequently, we suggest exploiting the RTMs to systematically test the adaptation engine and its parts in form of one-way testing of individual steps and fragments, in-the-loop testing of the analysis and planning steps, and online testing of the analysis and planning steps. 
We further study how we can validate the model which is required for the in-the-loop testing.

\section{One-Way Testing}
\label{sec:one-way-testing}
\noindent
We define \emph{One-Way Testing} as the following: An input RTM and an expected oracle RTM are provided. One or more steps are tested in a single execution of a partial feedback loop. The tested parts receive the input RTM and are supposed to produce an output RTM. The output RTM is compared against the oracle. In this kind of testing the steps $\to_{AS}$, $\to_{ENV}$, $\to_{M}$, $\to_{A}$, $\to_{P}$ or $\to_{E}$ will happen at most once. 

\subsection{One-Way Testing single MAPE Activities}\noindent
The most basic approach is to test each of the steps/activities that process the RTM on their own. Obviously, these tests need to be run before testing  combinations of feedback-loop steps to better locate faults and tell single-step errors from errors that arise due to problems in the interaction of steps.
\subsubsection{One-Way Testing the Analysis}\noindent
If we want to test the analysis step, we simply provide an input RTM $S_1^M$, run step $\to_A$, and compare the resulting RTM $S_1^A$ with an oracle RTM $S_o^A$.
Applied to the example in Figure~\ref{fig:trace}, we choose $S_{i}^M$ with the removed component as an input RTM. We then define an oracle RTM $S_o^A$ that contains an annotation where the missing component has been marked. Applying $\to_A$ on $S_{i}^M$ would give us $S_{i}^A$ which is compared to  $S_o^A$. If both RTMs are the same, that is, both especially contain the same ``missing component'' annotation, the test would pass, otherwise fail.
\subsubsection{One-Way Testing the Planning}\noindent
Similar to the analysis step, we provide an input RTM $S_1^A$, run step $\to_P$, and check whether the ouput of $\to_P$ is equal to an oracle RTM $S_o^P$ that was defined before. 
In the example of Figure~\ref{fig:trace}, we start out with the annotated RTM $S_{i}^A$. The oracle $S_o^P$ would be defined as the intact architecture from the beginning ($S_{i-1}^M$) and we would expect $\to_P$ to return an RTM equal to $S_o^P$, that is, the plan step has re-created the removed component in the RTM.   

\subsection{One-Way Testing MAPE Fragments}
\noindent
We now discuss one-way testing of fragments by jointly testing the analyze and plan or the monitor and execute steps.

\subsubsection{One-Way Testing the Analysis and Planning}\noindent
As a precondition to the separate test of the analysis and planning, it is necessary to have knowledge about the way the analysis works and what kind of models to expect. Obviously it would be hard to create a valid oracle model $S_o^A$ or input model $S_1^M$ if this knowledge is not available. In a simple scenario like the self-healing one presented before this should not pose a problem. But there are also more complex analysis algorithms, which will not result in models that can be tested as easily. Furthermore, some errors might only appear if the analysis and planning are tested together.

Consequently, we propose to test the analyze and plan steps as the next unit. Again we can benefit from the same pattern of testing, that is, by providing an input model $S_1^M$ and an oracle model in state $S_o^P$. In terms of the example trace (Figure~\ref{fig:trace}), this means to start with the broken monitored input model $S_i^M$, construct an expected model $S_o^P$ where the removed component is redeployed and check whether the resulting model of the application of $S_i^M \to_A \to_P$ $S_i^P$ is equal~to~$S_o^P$.  

\subsubsection{One-Way Testing the Execute and Monitor}\noindent
The separate testing of the monitor and execute steps via the runtime models is not feasible as the effect of the execute step cannot be directly observed. If we follow the same pattern as with the analysis and planning, we would end up with no result model for the execute step and no input model for the monitor step. The effect of the execute step cannot be directly observed since it is part of the concrete adaptable software. Likewise, the monitor step's input is directly obtained from the software. 
Instead of the separate testing, we propose to test the monitor and execute steps together. In this setup we need a working adaptable software and the tested execute and monitor steps are effecting and sensing the software. The test input is provided by a model $S_1^P$ to the execute step $\to_E$ which will effect the adaptable software. The adaptable software is monitored $\to_M$ and a new runtime model is obtained $S_2^M$. 

Equality and inequality of these two models can be interpreted in different ways: (1) equal models may indicate that the monitor and execute steps work correctly, (2) equal models may also mean that a failure in the execute step is masked by a failure in the monitor step (or the other way round), or (3) that the adaptable software or the environment mask a fault of the execute and/or monitor steps. If $S_1^P$ and $S_2^M$ are not equal, then either (4) the execute step, (5) the monitor step or (6) both do not work properly or (7) the environment introduced an error or the adaptable software showed erroneous behavior. 

Cases (3) and (7) can be ruled out by applying the test several times. It is unlikely that the environment will introduce the same error for all test runs and if the adaptable software was tested before, it is equally unlikely that it will constantly show erroneous behavior. In the cases (4), (5) and (6) we can assume a broken monitor and/or execute step. Case (1) should be more likely than (2) since it is not impossible but hard to have two faults that mask each other. Case (2) should become less likely the more tests with different $S_1^P$ and $S_2^M$ are done. In the end, equal models are a good indicator of working execute and monitor steps and non-equal models show that at least one of them is broken. 

With this test setup, only parts of the monitoring capabilities can be tested since its purpose is to detect not only correct but also incorrect states of the adaptable software. On the other hand, the execute step is not intended to have an effect on the software that causes an incorrect state. Therefore, we need to be able to impose an ``incorrect'' RTM on the adaptable software (such as $S_i^M$ in Figure~\ref{fig:trace}), so that we can test whether the monitor step is able to properly observe this incorrect state and create the according RTM. A special test adapter is needed, so that first a correct RTM can be imposed by the execute step and then the incorrect parts are added by the test adapter. The incorrect input RTM $S_{1 err}$ needs to be split into $S_{1 valid}$ which will be provided to $\to_E$ and $S_{1 invalid}$ which is given to the test adapter. The oracle $S_o^M$ for this test looks like $S_{1 err}$ and the monitor should observe an incorrect RTM. 

\section{In-the-Loop Testing}
\label{sec:in-the-loop-testing}
\noindent
Considering the analysis and planning, one-way testing is effective to find errors that always show up, independent from their previous executions in the feedback loop. If we want to identify errors that arise from an accumulated state of the system, we need to test them with sequences of inputs. It would be a cumbersome task to construct these sequences by hand. Instead we propose to provide a simulation that captures the behavior of the adaptable software (AS), environment (ENV), monitor (M) and execute (E) steps. This simulation will provide sequences of RTMs to the analyze step and will read back the RTMs from the plan step. 
We define such a \textbf{r}un\textbf{t}ime \textbf{m}odel \textbf{s}imulation by an automaton ${RTMS} = (\mathcal{S}_{RTMS}, \to_{RTMS})$ that comprises the combined behavior of AS, ENV, M, and E. Note that ${RTMS}$ is a simulation for testing purposes. The provided input RTM and the way the simulation model reacts to the output of $\to_A$ and $\to_P$ are supposed to be realistic but not an exact replacement for the real AS, ENV, M and E. It also means that it may behave non-deterministically to reflect realistic AS and ENV and therefore involves some random component. 

In order to decide whether a test is successful, we also need an oracle. In the simplest case the oracle is given by a state property $\phi$ for the model. In more complex cases $\phi$ may be even a sequence property or ensemble property. With respect to our example, the oracle may be the sequence property that some architectural constraints for our RTM are only violated for at most $n$ subsequent states.  

\subsection{Black-Box In-the-Loop Testing of Analysis and Planning}\noindent
With ${RTMS}$ at hand we can test the feedback loop already in an early stage when neither the adaptable software or the monitor and execute steps are available or ready. The analyze and plan steps combined with ${RTMS}$ can be simulated together and produce observable sequences: $\to_{RTMS} S_1 \to_A S_1^A \to_P S_1^P \to_{RTMS} S_2 \dots$. From these we consider only the traces of states: $\pi = S_1; S_1^P; S_2 \dots$ and check whether $\pi \models \phi$ to ensure that $\to_A$ and $\to_P$ as a black box work as expected.

\subsection{Grey-Box In-the-Loop Testing of Analysis and Planning}\noindent
We can also aim for a better fault location if we consider the result of $\to_A$ (i.e., the analyze and plan steps as a grey box). The sequence, we would like to look at, is the following: $\to_{RTMS} S_1 \to_A S_1^A \to_P S_1^P \to_{RTMS} S_2 \dots$. Here we will inspect the trace $\pi' = S_1; S_1^A; S_1^P; S_2 \dots$. In order to test these traces, we need a property $\phi'$ that covers $S_i^A$ as well. We now require $\pi' \models \phi'$ to ensure that $\to_A$ and $\to_P$ work as expected.     
 
\section{Online Testing and Validation}
\label{sec:online-testing}
\noindent
In a later development stage we can reuse the simulation model ${RTMS}$ and the properties $\phi$ and $\phi'$ alongside the running system for online testing and validation. 

\subsection{Online Testing}\noindent
If $\to_A$ and $\to_P$ in the running system will expose $S_i^A$ and $S_i^P$ in the same way as in the development stage, we can check $\phi$ and $\phi'$ online or against a recorded trace. The simulation is simply replaced with the real system. Whether online or offline testing is to be preferred will depend on available resources on the system under test and the existence of logging facilities. Both approaches, black-box and grey-box testing, are applicable and can be carried out in the same way as with the simulation. 

The sequences will be $(\to_{AS} or \to_{ENV})^* \to_M S_1^M \to_A S_1^A \to_P S_1^P \to_E (\to_{AS} or \to_{ENV})^*; \to_M S_2^M \dots$ and the traces will be the exchanged RTMs:
$\pi = S_1^M; S_1^A; S_1^P; S_2^M \dots$%

\subsection{Validation}\noindent
The in-the-loop testing heavily depends on ${RTMS}$. If an error is detected during in-the-loop testing, it is likely that it is caused by an erroneous adaptation ($\to_A$, $\to_P$ or both). But the ${RTMS}$ itself might also be the source of an error or might mask an erroneous adaptation. The validation of ${RTMS}$ in this later stage can give an indication about the quality of ${RTMS}$ and therefore the suitability for testing. Additionally, if the real system produces sequences not covered by ${RTMS}$ which cause errors in the adaptation, we exactly know which sequence reveals the error and it can be added to ${RTMS}$ for regression tests. 

The idea behind validating ${RTMS}$ is to observe $(\to_{AS} or \to_{ENV})^* \to_M S_1^M \to_A S_1^A \to_P S_1^P \to_E (\to_{AS} or \to_{ENV})^*; \to_M S_2^M \dots$ and look at the traces $\pi' = S_1^M; S_1^P; S_2^M \dots$. If our simulation model ${RTMS}$ is correct, it should cover the observed behavior: $\pi' \in \mathcal{L}({RTMS})$.

\section{Initial Evaluation}
\label{sec:evaluation}
\noindent
In this section, we report on our initial evaluation of the testing scheme for SASS we are proposing in this paper. This evaluation shows the benefits of using (architectural) runtime models with respect to implementing a test framework by means of reusing MDE techniques. Moreover, it gives us preliminary confidence about the effectivity of the scheme when developing feedback loops. 

\subsection{One-Way Testing}
\noindent
To realize one-way testing, we developed a generic test adapter that loads the input model, triggers the adaptation steps such as analysis and planning to be tested, and finally, compares the resulting model with the oracle model. Developing such a test adapter has been simplified due to MDE principles as realized by the \textit{Eclipse Modeling Framework} (EMF)\footnote{\textit{EMF}: \url{https://eclipse.org/modeling/emf/}}. EMF provides mechanisms to generically load and process models and particularly of comparing models\footnote{\textit{EMF Compare}: \url{https://www.eclipse.org/emf/compare/}}. Hence, we easily obtain matches and differences between two models such as the output model of adaptation steps and the oracle model to obtain the testing result. This result, that is, the output of the comparison, is also a model that can be further analyzed. For instance, the \textit{Object Constraint Language} (OCL)\footnote{\textit{Eclipse OCL}: \url{http://projects.eclipse.org/projects/modeling.mdt.ocl}} can be used to check application-specific constraints such as mission-critical components like for authenticating users on the mRUBiS marketplace are not missing in the architecture.

\subsection{In-the-Loop Testing}
\noindent
For the internet marketplace mRUBiS we developed a simulator based on an architectural runtime model. It simulates the marketplace itself (i.e., the adaptable software) thereby injecting failures as well as the monitor and execute steps. The simulator maintains the runtime model against which the analyze and plan steps are developed. 

Using this simulator, we can test the analyze and plan steps as follows: 
\begin{inparaenum}[i)]%
\item the simulator injects failures into the runtime model (this simulates the behavior of the adaptable software and environment as well as the monitor step that reflects the failure in the model).
\item the analyze and plan steps to be tested are executed and they analyze and adjust the model according to the adaptation need.
\item the simulator performs the execute step that emulates the effects of the adaptation as performed by the analyze and plan steps in the runtime model. For instance, response times are updated in the model if the configuration of the architecture is adapted.
\end{inparaenum}

After one run of the feedback loop and before injecting the next failures, the simulator checks whether the analyze and plan steps performed a well-defined adaptation (e.g., by checking whether the life cycle of components has not been violated when adding or removing components) and it checks whether the state of the runtime model represents a valid architecture (e.g., components are not missing or there are no unsatisfied required interfaces, that is, no dangling edges). 
These checks are performed based on constraints and properties that the runtime models must fulfill and the results of these checks are given as feedback to the engineer.

This simulator has been used in research and in courses to let students develop and test different adaptation techniques (e.g., hard-coded event-condition-action, graph transformation, or event-driven rules) for the analyze and plan steps. Though the simulator helped in finding faults in the adaptation logic, the randomness included in the simulator and its basic logging facilities impeded the automated reproducibility of traces and therefore, the retesting of ``interesting'' edge cases.

\subsection{Online Testing and Validation}
\noindent
So far, we have not worked on testing the adaptation online. However, our experience with runtime models and employing MDE techniques at runtime for self-adaptation~\cite{VG-TAAS-EUREMA,Vogel-ICAC09} gives us promising confidence to achieve the online testing. For instance, our EUREMA interpreter~\cite{VG-TAAS-EUREMA} that executes feedback loops already maintains the runtime models used within the loops and passes them along the loop's adaptation activities. Thus, when passing models along the activities, the interpreter may defer the execution of the next activity.
Before proceeding, the interpreter can either (1) hand over to an online testing activity that will compare the current RTM to one which is derived from a simulation model that runs in parallel or (2) log the RTM for later comparison in a simulation module. As discussed earlier, (1) has the advantage of immediate revelation of errors but needs computation resources on the system while (2) can benefit from more resources offline but needs persistance resources for the logs. In both cases, working only on changes of the RTM might reduce cost.     

\section{Related Work}
\label{sec:related-work}
\noindent
Testing of SASS has been addressed by others as well. This related work could usually be assigned to one of the following categories:
\begin{inparaenum}
\item \label{inpar:formal} The adaptation is formally specified and verified with special constructs regarding the adaptation~\cite{Sama2008, Iftikhar2012}, 
\item \label{inpar:online} the SASS is tested/verified at runtime/online and the verification expressions are adapted to properties unique to adaptation~\cite{Goldsby2008, Zhao2006, Eberhardinger2014}, 
\item \label{inpar:evolved} tests are evolved at runtime in an attempt to test for requirement fulfillment even when the environment or the adaptable software changes~\cite{Fredericks2015,Fredericks2014a}, and
\item \label{inpar:destime} testing is carried out at design time addressing the special issues of adaptive systems~\cite{Pueschel2014,Wang2007,Camara2014}.
\item \label{inpar:combined} Combined assessment of quality assurance for self-adaptive systems from more than one direction has also been done~\cite{Weyns2012}. 
\end{inparaenum} 
 
The work presented in~\cite{Weyns2012} already shows that a single quality assurance technique is not enough as an adequate approach to achieve high-quality SASS. Testing and formal verification have long been known as complementing techniques for most kinds of systems. We assume that quality assurance for SASS can benefit in the same way from the combination of approaches like the one presented by us and approaches of category \ref{inpar:formal}. Likewise, we see early testing as a complementary technique to online and adaptive online testing (cf. categories \ref{inpar:online} and \ref{inpar:evolved}). To our understanding, this specifically holds for SASS where unknown circumstances may arise at runtime and need to be adequately taken care of. Nevertheless, testing still needs to be done before a system is to be deployed to ensure at least an initial and basic quality of the SASS. 

Approaches of category \ref{inpar:destime} also address testing of SASS at design time. We differ from these approaches by not being dependent on a complete system. Using RTM as the test interface allows us to test already when there are only fragments of the system available, which is in the earlier development stages. Also our approach allows to test in a bottom-up manner, starting from the smallest testable units of a SASS and proceeding to the entire system. 
  
\section{Conclusion}
\label{sec:conclusion}
\noindent
In this paper we presented a systematic testing scheme for SASS. It encompasses a staged testing process inspired by the engineering of embedded software. Exploiting architectural runtime models with their various states allows us to address the different stages of one-way, in-the-loop, and online testing. Supporting early development stages with tests, we may find errors early. Furthermore, looking at the individual MAPE-K activities and their different integrations, we should be able to locate faults more easily. In this context, our initial evaluation gives us preliminary confidence about the scheme's effectivity.

There are several directions to evolve the presented testing scheme in future. 
As of now we employ an ad hoc simulator for ${RTMS}$. We could instead make use of a formal model to automatically derive test cases by using coverage criteria, which includes the generation of test inputs and oracles (runtime models and properties) taking the uncertainty of SASS and its environment into account.
Useful formalisms range from simple finite state machines to timed, hybrid or even probabilistic automata. 
Such a formal approach will further ease a thorough evaluation of the testing scheme.
Another direction would be to address the neglected case that the adaptable software and environment change while the MAPE loop is running. We will study special test setups for this case.

\bibliographystyle{IEEEtran}
\bibliography{references}

\end{document}